\documentstyle[preprint,aps,amssymb,psfig]{revtex}  
\begin{document}
\draft                       
\tighten


\title{Channel diffusion of sodium in a silicate glass}                           
\author{Philippe Jund, Walter Kob, and R\'emi Jullien}

\address{Laboratoire des Verres - Universit\'e Montpellier 2\\
         Place E. Bataillon Case 069, 34095 Montpellier France
	}

\date{June 11, 2001}

\maketitle


\begin{abstract}
We use classical molecular dynamics simulations to study the dynamics
of sodium atoms in amorphous Na$_2$O-4SiO$_2$. We find that the sodium
trajectories form a well connected network of pockets and channels. Inside
these channels the motion of the atoms is not cooperative but rather
given by independent thermally activated hops of individual atoms between
the pockets. By determining the probability that an atom returns to a
given starting site, we show that such events are not important for the
dynamics of this system.

\end{abstract}



The question of ionic transport in glasses has been the topic of
research for quite some time since it is linked to important
transport phenomena such as electrical conductivity, ion exchange, or
aqueous corrosion~\cite{ingramr,ngai,day}. Obviously the general problem
of the motion of the ions inside the glass is complex due to the inherent
structural disorder of the amorphous matrix. Consequently the nature of
the dynamics on short length scales, i.e. the local motion of the ions,
as well as the one on large scales, i.e. the global properties of the
trajectories, is understood only poorly.  Concerning the local motion
of the ions, models based on a ``forward-backward hopping mechanism''
\cite{funke} or hopping processes over random potential barriers
\cite{roling} have been proposed. Furthermore it has been suggested that
the combination of the disorder and the long-ranged Coulomb interactions
between the ions gives rise to a strong backward correlation of the motion
and is the reason for the Non-Debye behavior observed in the dielectric
response of a variety of materials~\cite{maas}.  Regarding the dynamics
on large scales, a picture based on ``preferential pathways'' proposed
some ten years ago~\cite{ingram} is quite popular~\cite{kahnt}. In that
direction, recent molecular-dynamics simulations of silica glasses with
different concentrations of sodium~\cite{oviedo} have given some support
to the ideas put forward by Greaves that ``the clustering of alkalis in
oxide glasses marks out the pathways for ionic diffusion''~\cite{greaves}
if the concentration of alkalis is higher than a percolation limit
composition.  However, as stated by the authors of Ref.~\cite{oviedo}:
``these suggestions are, as yet, unconfirmed''.  In this paper we
therefore present the results of a classical molecular-dynamics simulation of
Na$_2$O-4SiO$_2$ [NS4]. In particular we investigate in detail the
dynamics of the alkali ions on short as well as long length scales to
check the validity of some of the ideas described above.

Our system consists of $N_{\rm Na}$=86 sodium, 173 silicon and 389
oxygen atoms confined in a cubic box of edge length $L=20.88$~\AA,
which corresponds to the experimental value of the density
(2.38~g/cm$^3$)~\cite{bansal_86}. The interaction between the ions
is given by a modified version~\cite{wk} of the potential proposed by
Kramer {\em et al.}~\cite{kramer}. The potential $U(r_{ij})$ between two
particles $i$ and $j$ includes a Coulombic part, and a short range part 
and is given by 

\begin{equation}
U(r_{ij}) = \frac{q_iq_je^2}{r_{ij}} + A_{ij}\exp(-B_{ij}r_{ij})
            - \frac{C_{ij}}{r_{ij}^6} + \frac{\tilde{q}_i\tilde{q}_je^2}{r_{ij}}\left[ 1- \left( 1-\delta_{i{\rm Na}} \right) ( 1-\delta_{j{\rm Na}}) \right]\Theta(r_{{\rm c}}-r_{ij})
\end{equation}
with $q_{{\rm Si}}=\tilde{q}_{{\rm Si}}=2.4$, $q_{{\rm O}}=\tilde{q}_{{\rm O}}=-1.2$, $q_{{\rm Na}} = 0.6$ and $\tilde{q}_{{\rm Na}} = 0.6 \ln \left[ C \left(r_{{\rm c}}-r_{ij} \right)^2+1 \right]$ ($\Theta$ is the usual Heaviside function). More details and in particular the values of the parameters can be found 
in Ref.~\cite{wk}.  In previous studies
of the mixtures Na$_2$O-2SiO$_2$ [NS2] and Na$_2$O-3SiO$_2$ [NS3],
this potential has been shown to reproduce reliably many structural and
dynamic properties of sodium silicate melts~\cite{wk}. Therefore it can
be expected that it is also able to describe the salient features of
the microscopic dynamics of the present system.

After having equilibrated the liquid at 4000K for 50000 time steps (35ps)
we have cooled the system to 300K, with a linear cooling schedule at
a quench rate of $2.3\times10^{14}$ Ks$^{-1}$.  During this quenching
process we have saved the configurations (positions and velocities) at
several temperatures ($T=$4000, 3000, 2500, 2000, 1500 and 300 K), and
subsequently used them as starting configurations of production runs of 1
million steps (with a time step of 1.4 fs), giving thus for each temperature 
a total production time of 1.4 ns. Similarly to a previous study on pure
silica \cite{phmaga} a plot of the potential energy vs $T$ shows a change of 
slope around 2400~K which gives 
a rough estimate of the glass transition temperature in our system. 
In order to improve the statistics we have used, at 2000~K and 1500~K, three 
and two independent samples, respectively.

We first study the properties of the Na-trajectories at {\it long}
times. In Ref.~\cite{wk} it was shown that, for NS2 and NS3, the diffusion
constant for the sodium atoms is around two orders of magnitude higher
than the one of the silicon and oxygen atoms, if $T \leq 2500$K. Hence
it was concluded, in qualitative agreement with experiment, that the 
sodium atoms are diffusing inside a frozen matrix
of SiO$_2$. Although the atoms of this matrix vibrate around an equilibrium
position they do not show a structural relaxation on the time scale of
many ns. In the present work we therefore study the details of 
the dynamics of the Na atoms through this frozen matrix.  For this we 
investigate the space dependence of the distribution of the number 
density of the sodium atoms. To determine this distribution we divided 
the simulation box in $N_{\rm tot}=20^3$ distinct cubes (hence each has a 
volume of $\approx$ 1~\AA$^3$) and determined the number density of the 
sodium atoms in each of these small boxes. Thus this discrete distribution 
is a coarse grained approximation to the continuous one, the latter one 
being not accessible in a simulation. By varying the size of this mesh 
between 0.5~\AA\ and 2~\AA\ we made sure that the results obtained do not 
depend qualitatively on this size.

\begin{figure}
\psfig{file=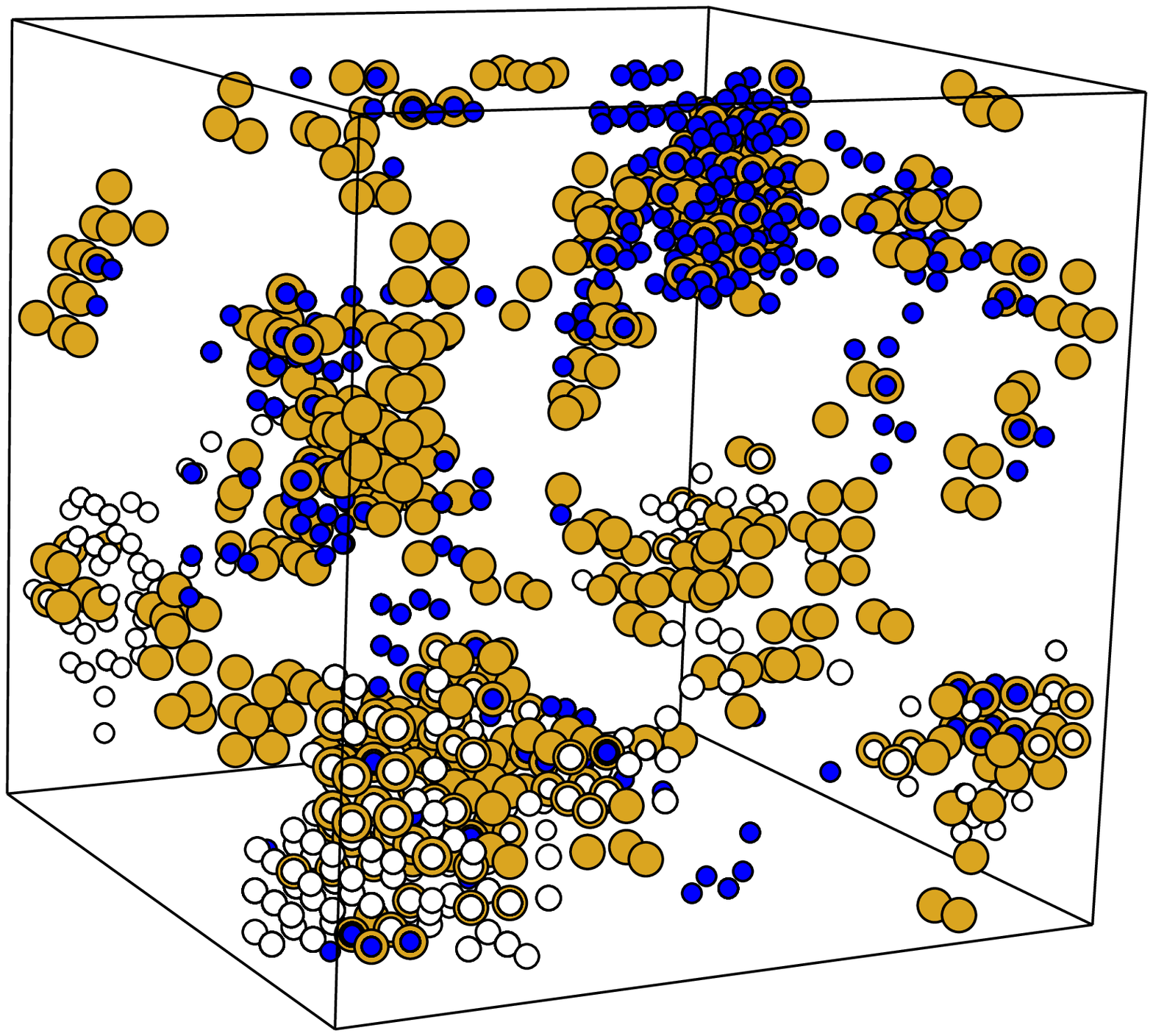,width=10cm,height=9cm}
\caption{
Snapshot of the simulation box. The grey spheres 
represent the
regions were at least seven different sodium atoms have passed during a 
1.4 ns run at 2000 K. The black
and white spheres show the trajectory of two individual sodium atoms.}
\label{fig1}
\end{figure}

In Fig.~\ref{fig1} we show those boxes which have been visited by at
least $\xi=11$ {\it different} Na atoms during a 1.4ns run at 2000~K
(grey spheres). The number of these boxes represents approximately 10\% of 
all the boxes visited by at least one sodium atom.
Therefore these spheres line out that region of space in 
which relatively many {\it different} sodium atoms passed. (Note that it 
is important to consider only different Na atoms; else one might also take into
account regions in which one sodium atom just oscillates around its
equilibrium position.)  From this figure we recognize immediately that
most of the Na-motion occurs in a relatively small subset ($\approx$ 6\%) of 
the total available space, i.e. that their trajectories do not fill the space
uniformly. This subset is itself composed of several blobs which are
connected to each other by rather loose structures. These pockets have
a typical size of around 3-6~\AA, but also larger ones can be found. This
size has to be compared with the typical cage which is seen by a sodium
atom on the time scale of the $\beta-$relaxation and which is on the
order of 1~\AA~\cite{wk}. Hence we conclude that these pockets are
somewhat larger than the size of the local cages in which the atoms rattle
back and forth at short times. The distance between these pockets is
typically 5-8~\AA, which has to be compared with the typical distance
between two sodium atoms (3.4~\AA, see Fig.~\ref{fig4} below) and an
experimental estimate for this quantity (1-2~\AA) \cite{sb}. 
(Note that this estimate has been obtained by using a very crude model, under 
the ``strong electrolyte assumption'' i.e. all the charge carriers are mobile,
 and neglecting cross correlations between the ions. If some of the 
assumptions of this model are changed slightly, an estimate for the distance
of 5~\AA\ is obtained \cite{sb}, in reasonable good agreement with the value 
found here). In order to check for finite size effects, we analyzed a NS3
sample containing 8192 atoms and found basically the same characteristic
dimensions of the blobs and connecting structures. We also mention that 
from Fig.~\ref{fig1} we see that these pockets are relatively well 
connected, which, as we will show below, is also reflected in the dynamics 
of the sodium atoms.

\begin{figure}
\psfig{file=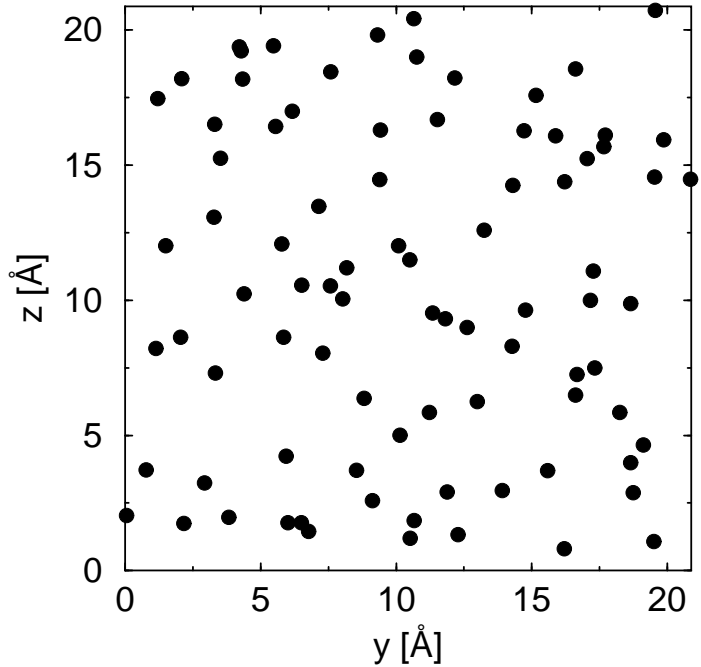,width=9cm,height=9cm}
\caption{Single snapshot of the positions of the Na atoms in the simulation 
box along the $x$-direction (similar snapshots are obtained along the $y$ 
and $z$ directions) after 700 ps at 2000K.
}
\label{fig2}
\end{figure}

Note that the result presented in Fig.~\ref{fig1} is
typical in the sense that other samples show a qualitatively similar
distribution of the density, although of course all the details are
different. Hence we confirm the idea proposed earlier~\cite{ingram} that
the trajectories of the sodium atoms lie in channels that go through the
SiO$_2$ matrix. We emphasize, however, that the structure
that we see in Fig.~\ref{fig1} is a {\it ``dynamic''} one in the sense
that it can only be found by averaging over the trajectories of all the Na
atoms and over a {\it sufficiently} long time. This is thus in contrast
to a {\it static} picture in which at {\it any} instant the sodium
atoms form some sort of percolating cluster~\cite{roling}. Therefore
a typical {\it single} snapshot of all the sodium atoms in the system, 
like the one presented in Fig.~\ref{fig2}, 
does not show any significant indication for the presence of these 
channels~\cite{sunyer}. We also mention that we find a strong 
correlation between the location of these pockets and the location of 
the non-bridging oxygen (NBO) atoms, which thus relates these pockets to the 
structure of the silica matrix~\cite{oviedo,sunyer}. When comparing the
mean square displacement of the NBOs with the one of the bridging oxygens,
we find that they are very similar and therefore that all the oxygen 
atoms are in an almost frozen state at 2000K. 

Also shown in Fig.~\ref{fig1} are the trajectories of two sodium
atoms (black spheres: the most mobile Na atom; white spheres: an arbitrary 
Na atom chosen at random). From these trajectories we see that individual 
atoms explore a substantial fraction of the channels. Hence we
conclude that the structure seen in the figure is not just an agglomerate
of the trajectory of many different atoms, but that most atoms visit a
large fraction of this structure.

The above chosen value of $\xi=11$ used to define the
channels is somewhat arbitrary in that decreasing or increasing $\xi$
will make the channels broader or thinner, respectively. Furthermore it
is clear that if one increases sufficiently the time over which the
density distribution is measured, at the end the Na atoms will have
visited all the boxes at least a few times, since even energetically very
unfavorable configurations will have a non-zero Boltzmann weight. Thus if
one wants to investigate the properties of the channels in more detail
it is necessary to determine where they have the {\it largest} Boltzmann
weight, since this will correspond to their core. This can be done as
follows: For every time $t$ we measure $n_i(t)$, the number of different
sodium atoms that have visited box $i$ in the time $t$ since the start
of the production run. Using $n_i(t)$ we therefore can define for every
time $t$ a distribution $P(k)$, which is the number of boxes that have
been visited exactly $k$ times. (Note that for short times the maximum
of $P(k)$ is at $k=0$, since most of the boxes have not been visited at
all). We now search for the boxes with high $n_i(t)$ i.e. those who have
been visited most frequently and hence contribute
to the tail of $P(k)$. For this we define the quantity $\xi(t)$ as the
largest integer such that the sum $N(t)=\sum_{k=\xi(t)}^{\infty}P(k)$ is
at least 10\% of the total sum, i.e. of $\sum_{k=1}^{\infty}P(k)$ (with this
definition, one recovers the value of $\xi=11$ used in Fig.~\ref{fig1}). Hence
 the boxes whose value of $n_i(t)$ is larger than $\xi(t)$ will be at
the core of the channels and therefore can be used to study them in more
detail. 

\begin{figure}
\psfig{file=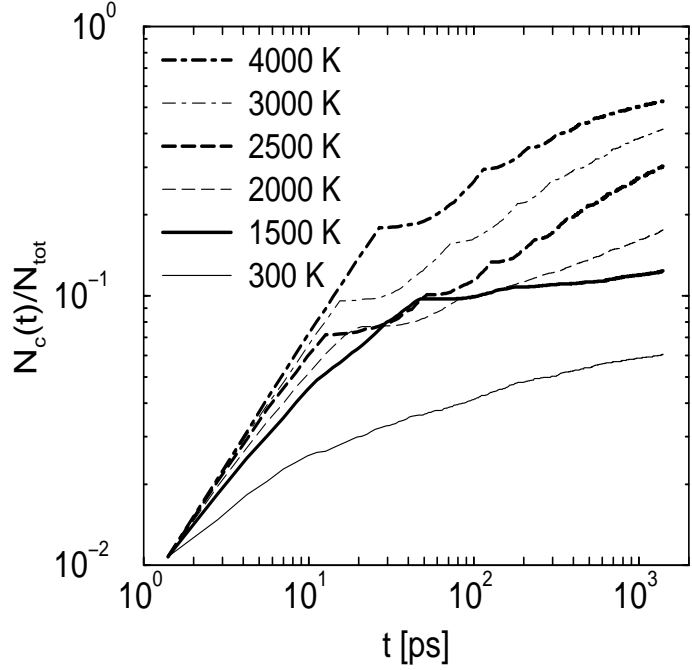,width=9cm,height=9cm}
\caption{
Time dependence of the fraction of cubes in the core (see text for
definition) of the channels for different temperatures.
}
\label{fig3}
\end{figure}

Equipped with these definitions we can now investigate the important question 
of how the structure of the channels depends on time, i.e. whether with 
time the sodium atoms explore more and more space in the matrix 
(by finding new pockets or pathways between pockets) or whether after a 
relatively short time they have explored most of the available space. To address this point, we define $N_c(t)$ which is the number of {\em different} boxes 
that have been (at any time) at the core of the channels 
(as defined above) after a time 
$t$ since the beginning of the simulation. In Fig.~\ref{fig3} we show how 
the fraction $N_c(t)/N_{tot}$ evolves with increasing time for different 
temperatures. At very short times ($t < 10$ps) this fraction 
increases rapidly with $t$. Then this increase slows down and we see that 
for $T=1500$~K the curve becomes relatively flat for times larger 
than around 50ps (a closer examination of the curve in this time regime 
shows that $N_c(t)$ increases logarithmically in $t$), which shows that at 
this temperature the channels are independent of time. Although at $T=2000$~K, the increase of $N_c(t)/N_{tot}$ after 50ps is
more important, the channels fill nevertheless only 17\% of the 
total available volume after 1.4ns. Therefore we conclude that at 
sufficiently low temperatures the structure of the channels is practically 
independent of time. We note a clear change of behavior for the curves
obtained at higher temperatures ($T \geq 2500$~K) both for the fraction
of space visited after 1.4ns (55\% at 4000 K) as well as for the increase at 
long time. This indicates that at these temperatures it is no longer 
reasonable to refer to channels since the sodium atoms can explore more 
and more new boxes as a consequence of the diffusing motion of the Si 
and O matrix atoms. A special case 
is the curve for $T=300$~K: For this low temperature the curve is still 
rising even at the end of the 1.4ns run. This indicates that the sodium atoms 
have not yet managed to explore the whole network of channels in this time. 
We have found that also this increase is proportional to log(t), with the 
{\em same} prefactor of the logarithm as the one found for the curve 
at $T=1500$~K at long times.
This shows that the mechanism how the sodium atoms explore the network is,
 at long times, independent of $T$, if the temperature is sufficiently low. 
Hence we conclude that the results studied here should also be valid in the
temperature range and time range probed in real experiments. Note that
this slow exploration of the network immediately rationalizes the 
experimental observation that the diffusion of the ions at low 
temperatures is anomalous and that the conductivity shows a dispersion. 

So far we have shown that at low $T$ the sodium atoms diffuse within a sub-space
of the total available space. Of interest is also how the Na atoms
move inside these channels. One possibility to address this point is
to investigate the distinct part of the van Hove function defined
by \cite{hmd}

\begin{equation}
G_d^{\rm NaNa}(r,t) = 
(L^3/4\pi r^2N_{\rm Na}^2)\sum_{i=1}^{N_{\rm Na}}\sum_{j=1}^{N_{\rm Na}}
{^{^\prime}}\langle\delta(r-|{\bf r}{_i}(t)-{\bf r}{_j}(0)|)\rangle .
\end{equation}
Thus, e.g., if at time $t=0$ there is a sodium atom at the origin,
the function $G_d^{\rm NaNa}(r=0,t)$ is proportional to the probability 
to find at time $t$ a {\it different} sodium atom at the origin. For $t=0$ this
function is proportional to the usual radial distribution function $g_{\rm
NaNa}(r)$.  In Fig.~\ref{fig4} we show $G^{\rm NaNa}_d(r,t)$ for various
times. We see that with increasing time the correlation hole at $r=0$
is quickly filled up and that a large peak grows at this distance. The
height of this peak attains a maximum after a time $\tau_{\rm max}$
and then decreases again. Hence we conclude that once a sodium particle
has moved from its location it had at $t=0$, there is a very high 
probability that a different Na atom
will take up its place and that this probability is highest after
the time $\tau_{\rm max}(T)$. In the inset of Fig.~\ref{fig4} we show
$G_d^{\rm NaNa}(r,\tau_{\rm max}(T))$ for different temperatures and
we recognize that the maximum height of the peak increases quickly
with decreasing temperature. This increase shows that with decreasing
$T$ the matrix becomes more and more rigid and hence maintains the
memory of the locations of the sodium atoms, even if they have moved
away since a long time.  The time $\tau_{\rm max}$ shows an Arrhenius
dependence with an activation energy around 1.3~eV. This value
is very well in line with those found in the study of the diffusion
coefficient of NS2 (0.93~eV) and NS3 (1.26~eV)~\cite{wk}. The results
shown in Fig.~\ref{fig4} indicate that the sodium atoms move by thermally
activated jumps between sites previously occupied by other sodium atoms,
in agreement with the proposition of Greaves and Ngai~\cite{ngai}. We
emphasize, however, that $\tau_{\rm max}(T)$ is {\it not} the time that 
an atom takes between a jump from one site to a neighbor site, which thus 
is related to the diffusion constant, but the time it takes until a site 
that has been freed is occupied by a new Na atom (which is not necessarily 
a nearest neighbor).

Note, however, that this result does {\it not} necessarily imply that 
the motion of the sodium atoms is collective. To investigate this point 
we have calculated how the distance between two sodium atoms 
changes with time. 
If the motion of the atoms is collective, it can be expected that
the increase of this distance $\Delta(t,\delta_0)$ is slower if $\delta_0$,
the initial separation between the two atoms, is small, than if
$\delta_0$ is large. In Fig.~\ref{fig5} we show the time dependence of
$\Delta^2(t,\delta_0)$ for sodium pairs that at time zero were nearest,
next nearest, or third nearest neighbors (bottom to top). (For this
classification we used the radial distribution function shown in 
Fig.~\ref{fig4}.) From this
figure we recognize that all three curves converge to the long time
limit $L^2/4$ with the {\it same} time constant. Thus since there
is no dependence on $\delta_0$ we conclude that the motion of the
atoms is {\it not} cooperative, in contrast to what has been suggested in
the literature~\cite{kahnt} and in agreement with (indirect) experimental
evidence \cite{sb}. In view of the fact that the snapshot in
Fig.~\ref{fig1} shows that each pocket has several connections (pathways)
to other pockets, such a quick decorrelation is, {\em a posteriori}, not
surprising~\cite{footnote}.

We also note that the decay of the correlation can be approximated well by
an exponential function with a time constant $\tau_\Delta$ independent of
$\delta_0$. In the inset
of Fig.~\ref{fig5} we plot $\Delta^2(t,\delta_0)$ for the first neighbor
shell, for all temperatures investigated, versus $t/\tau_\Delta(T)$. Since,
within the accuracy of the data, at low $T$ these curves collapse onto
a master curve, we conclude that the mechanism for the decorrelation,
and hence the motion, is independent of $T$. Finally we mention, that
also $\tau_\Delta(T)$ shows an Arrhenius dependence with an activation
energy around 1.3~eV, in agreement with the other time scales of the
sodium motion.

Certain models of the sodium dynamics assume that after a jump of an
atom, there is an enhanced probability that it jumps back to its original
site~\cite{funke}.  Therefore we have calculated the probability that a
sodium atom that was at the origin at time zero, moves for $t>0$ over a
distance which is larger than the beginning of its first nearest neighbor
shell (2~\AA, determined from $g_{\rm NaNa}(r)$) and then back to its
original site. We have found that the maximum value for this probability
is, at 2000K, around 26\%. This value has to be compared with the
``trivial'' one which is obtained if one considers that in this system
each sodium atom has around 5 other sodium atoms as nearest neighbors,
thus giving a probability to jump back of 1/5=20\%. Hence we conclude
that the atoms do not show a significant trend to jump back (backward
correlations), at least not in the temperature range studied here.

In summary we have shown that at low temperatures the long time
trajectories of the sodium atoms in the silica matrix occupy only a
relatively small fraction of the total space. This subset forms a well
connected network of channels and pockets which is explored very quickly
by each sodium atom.  We emphasize that from individual snapshots the
existence of these channels can hardly be seen~\cite{sunyer}, at least
for the sodium concentration studied here, which is in contrast to the
popular picture proposed by Greaves~\cite{greaves}.  The motion inside
these channels is not cooperative since nearby atoms decorrelate quickly
and this is also in distinction to common belief.  Furthermore we find 
that in this system the ``forward-backward jumps'' are not important. However,
it cannot be excluded that at higher concentrations of sodium such a
mechanism becomes relevant.

Part of the numerical calculations were done at CINES in
Montpellier.


%
\newpage

\begin{figure}
\psfig{file=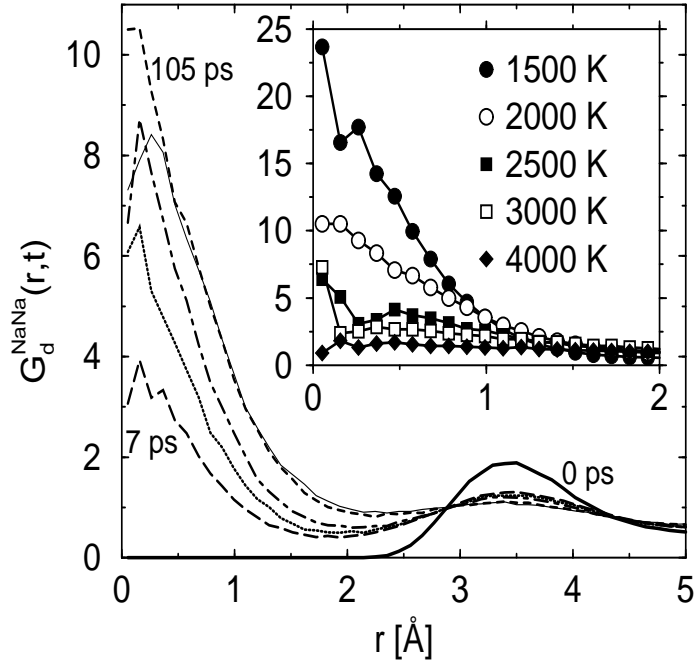,width=9cm,height=9cm}
\caption{
Plot of the distinct part of the van Hove function at 2000K for $t=0$
(solid bold), 7ps (long dashed), 28ps (dotted), 70ps (dot-dashed), 
105ps (dashed), and 140ps (thin solid). Inset: $G_d^{\rm NaNa}(r,\tau_{\rm
max}(T))$ at different temperatures, where $\tau_{\rm max}(T)$ corresponds 
to the time for which $G_d^{\rm NaNa}(0,t)$ is maximum.
}
\label{fig4}
\end{figure}

\begin{figure}
\psfig{file=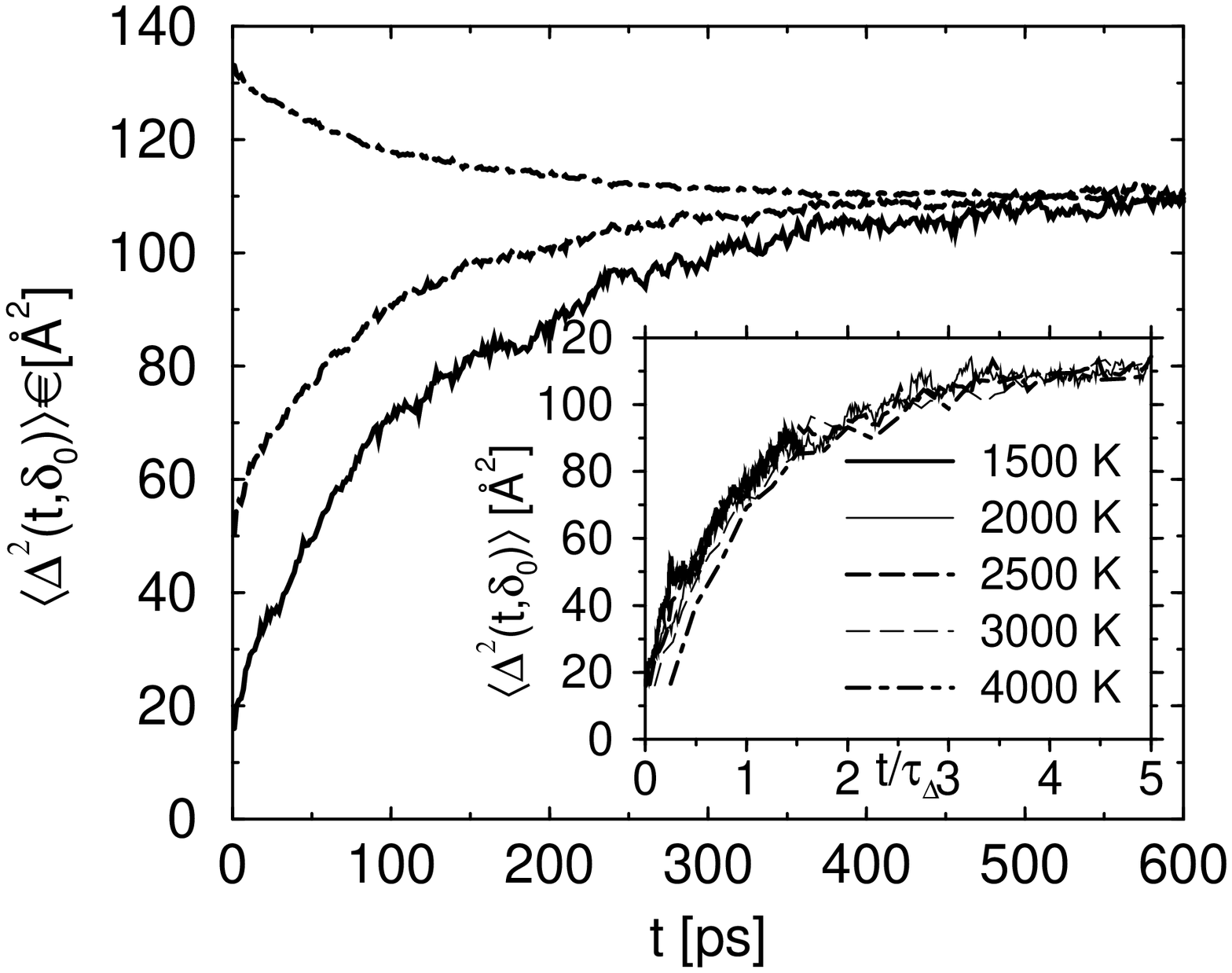,width=9cm,height=9cm}
\caption{
Time dependence of $\Delta^2(t,\delta_0)$, the squared distance
between two sodium atoms that at time zero were separated by a distance
$\delta_0$, at T=2000K. The three curves correspond to $\delta_0$ in the first,
second, third nearest neighbor shell (bottom to top). Inset:
$\Delta(t,\delta_0)$ for the first nearest neighbor shell, at
different temperatures, versus rescaled time.
}
\label{fig5}
\end{figure}



\begin{references}
\bibitem{footnote}
Note that the connectivity between the pockets is even higher than
suggested in Fig.~1, since due to the periodic boundary conditions not
all connections can directly be seen.
\end{references}
\end{document}